# FPGA Based Efficient Multiplier For Image Processing Applications Using Recursive Error Free Mitchell Log Multiplier And KOM Architecture

Satish S Bhairannawar[1], Rathan R[2], Raja K B[2], Venugopal K R[3], L M Patnaik[4]

[1]Department of Electronics and Communication Engineering,
Dayanand Sagar College of Engineering, Bangalore, India.
[2]Department of Electronics and Communication Engineering,
University Visvesvaraya College of Engineering, Bangalore, India
[3]Principal, University Visvesvaraya College of Engineering,
Bangalore University, Bangalore, India
[4]Honorary Professor, Indian Institute of Science, Bangalore, India.

## ABSTRACT

*The Digital Image processing applications like medical imaging, satellite imaging, Biometric trait images etc., rely on multipliers to improve the quality of image. However, existing multiplication techniques introduce errors in the output with consumption of more time, hence error free high speed multipliers has to be designed. In this paper we propose FPGA based Recursive Error Free Mitchell Log Multiplier (REFMLM) for image Filters. The 2x2 error free Mitchell log multiplier is designed with zero error by introducing error correction term is used in higher order Karastuba-Ofman Multiplier (KOM) Architectures. The higher order KOM multipliers is decomposed into number of lower order multipliers using radix 2 till basic multiplier block of order 2x2 which is designed by error free Mitchell log multiplier. The 8x8 REFMLM is tested for Gaussian filter to remove noise in fingerprint image. The Multiplier is synthesized using Spartan 3 FPGA family device XC3S1500-5fg320. It is observed that the performance parameters such as area utilization, speed, error and PSNR are better in the case of proposed architecture compared to existing architectures.*

## 1. INTRODUCTION

The multipliers play an important role in Digital Signal Processing (DSP) applications, where complex computations are involved. The Conventional Multipliers requires $n$ steps to compute the product. Each step having two parts i.e., in first part LSB of multiplier is verified in each step, if it is one, then multiplicand is added to partial product else zero is added to product. In Second part, multiplicand is shifted left and multiplier is shifted right, discarding the bit which has been shifted out. The conventional multiplier like array multiplier requires $(n-2)$ $n$-bits carry save adders and one *n* bit carry propagate adder for *n* bit multiplier, hence require more time, which decreases the speed of operation in complex DSP applications. In most DSP applications, speed is important criteria rather than accuracy, where truncated and Logarithmic multipliers are suited at most. In truncated multipliers, some of less significant partial products are discarded and compensation is provided, that partly compensates for left out terms. The Logarithmic multipliers is an alternate to fixed or floating point multipliers, where rounding of products are acceptable. The Logarithmic multipliers convert multiplication and division problem into addition and subtraction respectively. The focus of Log Multiplier is to reduce errors with power consumption. The LUT based Log multiplier consumes more area to store Log and Antilog values, where as the Mitchell log multiplier requires few shift operations for log and antilog calculations which require less hardware overhead, resulting in high speed and less power consumption.

*Contribution:* In this paper FPGA based Recursive Error Free Mitchell Log Multiplier for image filters is proposed. The error in 2x2 Mitchell log multiplier is corrected by introducing correction term. The higher order multipliers are derived from 2x2 error free multiplier using KOM parallel architecture.

*Organization:* The paper is organized as follows: Section 2 proposes the Literature survey followed by basic Mitchell multiplier algorithm and KOM. Section 3 gives details of Hardware design and implementation of proposed FPGA based REFMLM. Section 4 provides the proposed REFMLM algorithm. Section 5 proposes the performance analysis and results. Section 6 contains conclusions followed by future work.

## 2. RELATED WORK

Fhsfk Patricio Bulic et al., [1] proposed iterative logarithmic multiplier and compared different multipliers in a logarithmic number system to achieve an arbitrary accuracy through an iterative procedure. The error correction was performed in parallel with the basic multiplication. The limitation of this method is increase in combinational delay with each added correction circuit. Hall et al., [2] described a method for rapid multiplication with simple digital circuit. The algorithm consists of computing approximate binary logarithms, adding or subtracting the logarithms and computing the approximate antilogarithm of the result. The algorithm is used in parallel digital filter application but not simpler to compute single products compared to array multipliers. Abed and Sifred [3] proposed and implemented a unique 32-bit binary-to-binary logarithm converter CMOS. The converter calculates approximated logarithm in a single clock cycle and is implemented with pure combinational logic design. The fast Leading One Detector (LOD) circuit and modified logarithmic shifter used in the logarithmic converter operate at high speed and consume less area. The errors are introduced due to logarithmic approximations.

Abed and Sifred [4] presented VLSI implementation of a unique 32-bit antilogarithmic converter, which generates data for Digital Signal Processing (DSP) applications. This converter is implemented using combinational logic and computes antilogarithm in single clock cycle. McLaren [5] proposed a method to improve the accuracy of a logarithmic multiplier based on Mitchell's algorithms for calculating logarithms and antilogarithms. The error in the product using proposed technique is dependent only on the binary fraction components of two multiplicands which offers area and power saving. Davide De Caro et al., [6] proposed a technique based on mixed integer linear programming to obtain optimal co-efficient values that minimize the relative approximation error while using a reduced number of nonzero bits for the co-efficient. The hardware implementation realizes the multiplication by a few shifts and additions, avoiding the use of full multipliers. This technique yields complex hardware for low-precision implementations.

Ramin Tajallipour et al., [7] proposed an efficient algorithm for computing decimal logarithm using 64 bit floating point arithmetic. The algorithm is based on digit by digit iterative computation that does not require lookup tables, curve fitting, decimal to binary conversion and division algorithms. Bryson R Payne et al., [8] made comparisons between CPU's and GPU's. The research examines that image processing convolution are performed in better way compared to CPU. Muhammad Rais et al., [9] presented hardware design and implementation of FPGA based parallel architecture for standard and truncated multipliers. The proposed multiplier consumes less area compared to standard multipliers. Gang Zhou et al., [10] presented efficient FPGA implementations of bit parallel mixed Karatsuba–Ofman multipliers (KOM) and Galois Field GF ($2^m$). The common expression sharing and the complexity analysis on odd-term polynomials were introduced to achieve a lower gate bound than previous Application Specific Integrated Circuit (ASIC) discussions. The analysis was extended by using 4 input or 6 input lookup tables on FPGAs. Kowada et al., [11] proposed reversible circuits for Karastuba's algorithm to analyze computational complexity. The garbage disposal methods were discussed and compared with Bennet's schemes. This circuits can be used in reversible computers which have the advantage of being very efficient in terms of energy consumption.

Chip-Hong Chang and Ravi Kumar Satzoda [12] proposed an Adaptive Pseudo-carry Compensation Truncation (PCT) scheme which yields low average error among existing truncation methods. The proposed method achieves lower average. Atsushi Miyamoto et al., [13] proposed a systemic design approach to provide the optimized Rivest-Shamir-Adleman (RSA) processors based on high-radix Montgomery multipliers satisfying various requirements such as circuit area, operating time and resistance against side channel attacks. The proposed approach provides optimized data paths satisfying the the requirements by combining three data path architectures using different intermediate forms. The multiplier not best suited for cryptographic algorithms.

## 2.1 Mitchell Log Multiplier (MLM)

The MLM [14] is a Logarithmic Number System (LNS) multiplier which suits very well for applications where multiplication accuracy is not much of importance. In LNS multiplier, two operands are multiplied by finding their logarithms, adding the logarithm values and then calculating the antilogarithm of the sum to get required product of multiplier. The MLM generates approximated values of log and antilog values and are derived from following equations.

Consider a binary number N in the interval $2^{k+1} > N \geq 2^j$, where k is first MSB position one in binary and is considered as characteristic number and j is LSB position. The Binary number $N$ can be written as given in equations 1 and 2.

$$N \;=\; \sum_{i=j}^{k} 2^{\,i} \cdot z_i \qquad\qquad (1)$$

$$N = \sum_{i=j}^{k-1} 2^i z_i + \sum_{i=k} 2^i z_i,$$

$$= \sum_{i=j}^{k-1} 2^i z_i + 2^k z_k, \qquad \text{Considering } z_k = 1,$$

$$N = 2^k + \sum_{i=j}^{k-1} 2^i z_i,$$

$$= 2^k (1 + 2^{-k} \sum_{i=j}^{k-1} 2^i z_i)$$

$$N = 2^k (1 + \sum_{i=j}^{k-1} 2^{i-k} \cdot z_i) \qquad (2)$$

Where $z_i$ is a bit value of the $i^{\text{th}}$ position. Logarithmic value of $N$ is given in equations 3 and 4.

$$\log_2(N) = \log_2(2^k (1 + \sum_{i=j}^{k-1} 2^{i-k} \cdot z_i)$$

$$= \log_2(2^k (1 + x)) \qquad (3)$$

Where $\sum_{i=j}^{k-1} 2^{i-k} \cdot z_i = x,$

$$\log_2(N) = k + \log_2(1 + x) \qquad (4)$$

Since k ≥ j, $x$ will fall in the range 0≤ $x$ <1. Where $x$ is the fraction or mantissa, and $j$ depends on the numbers precision (it is 0 for integer numbers). The logarithmic product of two binary numbers $N_1$ and $N_2$ is given in equation 5.

$$\log_2(N_1 \cdot N_2) = k_1 + k_2 + \log_2(1 + x_1) + \log_2(1 + x_2) \qquad (5)$$

($\log_2 (1 + x)$) is usually approximated as '$x$' as given in equation 6

$$\log_2(1 + x) = x \qquad (6)$$

Substitute (6) in (5)

$$\log_2(N_1 \cdot N_2) = k_1 + k_2 + x_1 + x_2 \qquad (7)$$

The logarithm of the product of two numbers is expressed as the sum of their characteristic numbers and mantissas as given in equation 7.

For 16-bit numbers, characteristic number k1 and k2 ranges from 0 to 15, while mantissas $x_1$ and $x_2$ range in between 0 and 1. The product of MLM ($P_{\text{MLM}}$) is obtained by applying antilog as given in equation 8.

$$P_{MLM} = 2^{k_1+k_2}(1+x_1+x_2), \quad for \ x_1+x_2 < 1$$
$$= 2^{k_1+k_2+1}(x_1+x_2), \quad for \ x_1+x_2 \geq 1$$

$$(8)$$

The final approximation for the product requires the comparison of the sum of the mantissas with *one*. The sum of the characteristic numbers determines the most significant bit of the product. The sum of the mantissas is then scaled (shifted left) by $2^{k_1+k_2}$ or $2^{k_1+k_2+1}$ depending on the value of $x_1+x_2$. If $x_1+x_2 < 1$, the sum of mantissas is added to the most significant bit of product to complete the final result otherwise, the product is approximated only with the scaled sum of mantissas. The Mitchell multiplier produces a significant error percentage due to approximations made. The relative error increases with the number of bits with the value of *one* in the mantissas. An example of binary multiplications of two numbers with non power of two operands using MLM is given in Figure 1. The binary multiplication value using MLM is differing from actual multiplication value resulting in error. An example of binary multiplication of two numbers with at least one number having power of 2 using MLM is given in Figure 2. It is observed from example given in Figure1 that MLM has errors for operands with non powers of two.

> Consider non powers of 2 operands to be multiplied using MLM.
>
> $18 = 00010010_{(2)}$; $60 = 00111100_{(2)}$
>
> Characteristic k1 of 18 = $100_{(2)}$; mantissa $x_1 = 0010_{(2)}$
>
> Log (18) = $k_1$. $x_1 = 100.0010$
>
> Characteristic $k_2$ of 5 = $101_{(2)}$; mantissa $x_2 = 11100_{(2)}$
>
> Log (60) = $k_2$. $x_2 = 101.11100_{(2)}$
>
> Log 18 + log 60 = 1010. 0000; $x_1 + x_2 > 1$
>
> $P_{MLM} = 2^{(k_1+k_2+1)}$ appended with $(x_1+x_2) = 2^{4+5+1}$ appended with 0000
>
> $P_{MLM} = 10000000000_{(2)} = 1024$
>
> $P_{TRUE} = 1080$
>
> Error = $P_{TRUE} - P_{MLM} = 1080 - 1024 = 56$

Figure1. Example of MLM with non powers of 2 operands.

## 2.2 Error Correction terms for Mitchell multiplier by Mitchell.

The error in Mitchell multiplier [14] is always positive so it can be reduced by successive multiplications by using analytical expression for the error correction. The Mitchell approximated product is given in equation 9.

$$P_{MLM} = 2^{k_1+k_2}(1+x_1+x_2), \quad for \ x_1+x_2 < 1$$
$$= 2^{k_1+k_2+1}(x_1+x_2), \quad for \ x_1+x_2 \geq 1$$

$$(9)$$

The true product without approximation for Mitchell Log multiplier is given in equation 10.

$$P_{true} = 2^{k_1}(1+x_1) \cdot 2^{k_2}(1+x_2),$$
$$= (2^{k_1}+2^{k_1}x_1) \cdot (2^{k_2}+2^{k_2}x_2),$$
$$= 2^{k_1+k_2}(1+x_1+x_2+x_1 \cdot x_2)$$

$$(10)$$

The error between true product and approximated product is given in equation 11 for condition $x_1 + x_2 < 1$

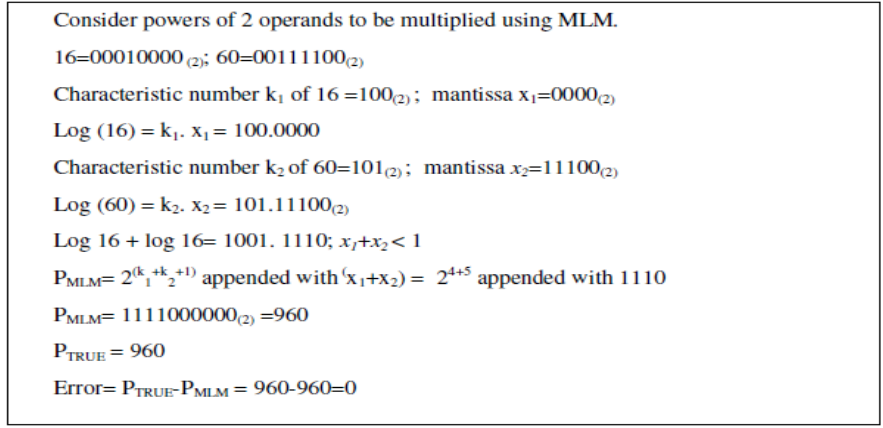

Consider powers of 2 operands to be multiplied using MLM.

$16 = 00010000_{(2)}$; $60 = 00111100_{(2)}$

Characteristic number $k_1$ of 16 = $100_{(2)}$; mantissa $x_1 = 0000_{(2)}$

Log (16) = $k_1$. $x_1$ = 100.0000

Characteristic number $k_2$ of 60 = $101_{(2)}$; mantissa $x_2 = 11100_{(2)}$

Log (60) = $k_2$. $x_2$ = 101.11100$_{(2)}$

Log 16 + log 16 = 1001. 1110; $x_1 + x_2 < 1$

$P_{MLM} = 2^{(k_1 + k_2 + 1)}$ appended with '$(x_1 + x_2)$ = $2^{4+5}$ appended with 1110

$P_{MLM}$ = 1111000000$_{(2)}$ =960

$P_{TRUE}$ = 960

Error= $P_{TRUE} - P_{MLM}$ = 960-960=0

Figure2. Example of MLM with one of the operand in powers of 2.

$$P_{true} - P_{MLM} = 2^{k_1 + k_2}(1 + x_1 + x_2 + x_1 \cdot x_2) - 2^{k_1 + k_2}(1 + x_1 + x_2),$$
$$= 2^{k_1 + k_2}(1 + x_1 + x_2 + x_1 \cdot x_2 - 1 + x_1 + x_2)$$
$$= 2^{k_1 + k_2}(x_1 \cdot x_2) \tag{11}$$

The error between true and approximated product for condition $x_1 + x_2 \geq 1$ is given in equation 12.

$$P_{true} - P_{MLM} = 2^{k_1 + k_2}(1 + x_1 + x_2 + x_1 \cdot x_2) - 2^{k_1 + k_2 + 1}(x_1 + x_2),$$
$$= 2^{k_1 + k_2}(1 + x_1 + x_2 + x_1 \cdot x_2 - 2(x_1 + x_2)) \tag{12}$$
$$= 2^{k_1 + k_2}(1 + x_1 \cdot x_2 - (x_1 + x_2))$$

The equation 12 can be simplified by letting $x_1 = 1 - x_1'$ and $x_2 = 1 - x_2'$, where $x_1'$ and $x_2'$, are the two's complement of $x_1$ and $x_2$ as given in equation 13.

$$P_{true} - P_{MLM} = 2^{k_1 + k_2}(x_1' \cdot x_2') \quad for \quad x_1 + x_2 \geq 1 \tag{13}$$

The error in multiplication can be eliminated if these correction terms are added into the $P_{MLM}$. The MLM with error correction is given using equation 14.

$$P_{MLMC} = P_{MLM} + 2^{k_1 + k_2}(x_1 \cdot x_2), \quad for \quad x_1 + x_2 < 1$$
$$= P_{MLM} + 2^{k_1 + k_2}(1 - x_1) \cdot (1 - x_2), \quad for \quad x_1 + x_2 \geq 1 \tag{14}$$

It is clearly observed from equation 14, that the corrected product consists of two extra multiplications, hence to correct error one more multiplier is required, which is again a disadvantage. So Mitchell corrections terms has drawback of successive multiplications.

**2.3 Karatsuba Ofman Multiplier (KOM)[15]**

The high speed multiplier which is used to multiply two large numbers by decomposing each numbers into two half sized numbers. General multiplication of two $n$ digit numbers need $n^2$ single digit products, where as KOM reduces at most to $3n^{(\log_2 3)} = 3n^{(1.585)}$ single digit multiplications and exactly $n^{(\log_2 3)}$ when n is a power of two. If $n = 2^{10} = 1024$, in particular, the exact counts are $(2^{10})^2 = 1{,}048{,}576$ for general multiplications and $3^{10} = 59{,}049$ for KOM.

Let $a$ and $b$ are two $n$-digit numbers to be multiplied using KOM are in radix 2 where $n$ is restricted even. Decompose $a$ and $b$ into two parts as given in equation 15 and 16 respectively.

$$a = a_L + a_H \, z^{n/2}, \tag{15}$$

$$b = b_L + b_H \, z^{n/2}, \tag{16}$$

Where $a_L$ = lower order bits varying from 0 to $n/2$-1,
 $\quad\quad$ $b_L$ = lower order bits varying from 0 to $n/2$-1,
 $\quad\quad$ $a_H$ = higher order bits varying from $n/2$ to $n$-1,and
 $\quad\quad$ $b_H$ = higher order bits varying from $n/2$ to $n$-1.

The product of $a$ and $b$ using KOM is given in equation 17

$$
\begin{aligned}
P_{KOM} &= a \cdot b \\
&= (a_L + a_H z^{n/2})(b_L + b_H z^{n/2}) \\
&= a_L b_L + (a_L b_H + a_H b_L) z^{n/2} + (a_H b_H) z^n
\end{aligned}
\tag{17}
$$

Thus, we can compute the product $P_{kom}$ from 4 half-sized products $a_L b_L$, $a_L b_H$, $a_H b_L$, and $a_H b_H$. Further these four half sized products can be reduced to three products at the cost of few more additions using equation 18.

$$a_L b_H + a_H b_L = a_L b_L + a_H b_H + (a_L - a_H)(b_H - b_L) \tag{18}$$

Substitute equation 17 in equation 18 to get optimized $P_{kom}$ given in equation 19.

$$
\begin{aligned}
P_{KOM} = a_L b_L &+ \{a_L b_L + a_H b_H + (a_L - a_H) \\
&(b_H - b_L)\} z^{n/2} + a_H b_H z^n
\end{aligned}
\tag{19}
$$

The multiplication of two operands in binary using KOM is described with an example in Figure 3. The product of two operands has no error.

# 3. PROPOSED FPGA BASED REFMLM FOR IMAGE FILTERS

In this section, we describe a new approach to design higher order multiplier using KOM with basic 2x2 multiplier developed by EFMLM by introducing correction term. Since the proposed method use 2x2 EFMLM as basic multiplier for high order multipliers, the average error rate is zero. The proposed method is a combination of KOM and EFMLM with radix 2 decomposition as shown Figure 4. The 16x16 multiplier to be implemented is decomposed into smaller order multipliers using radix2 concept i.e., 16x16 is decomposed into 8x8 (4 multipliers), 4x4 (16 multipliers) and 2x2 (64 Multipliers). The EFMLM is used in every 2x2 multiplication that results in zero error. The KOM is used in every 4x4, 8x8 and 16x16 multiplications along with 2x2 EFMLM to obtain error free multiplication values.

Let 16 and 60 be two operands to be multiplied.

$a$=16=00010000$_{(2)}$ & $b$=60=00111100$_{(2)}$; Here Z=2 since binary and $n$= number of Ptruebits= 8

$a_L$=0000$_{(2)}$, $a_H$=0001, $b_L$=1100$_{(2)}$ and $b_H$=0011$_{(2)}$

$a_L b_L$=00000000$_{(2)}$; $a_H b_H$=00000011$_{(2)}$

$(a_L - a_H)(b_H - b_L)$= (-0001$_{(2)}$)(-1001$_{(2)}$)=00001001$_{(2)}$;

$a_L b_L + a_H b_H + (a_L - a_H)(b_H - b_L)$=00001100$_{(2)}$

$P_{kom} = a_L b_L + \{a_L b_L + a_H b_H + (a_L - a_H)(b_H - b_L)\} z^{n/2} + a_H b_H z^n$

= 00000000$_{(2)}$+000011000000$_{(2)}$+0000001100000000$_{(2)}$

$P_{kom}$=1111000000$_{(2)}$ =960, $P_{true}$ =960

Error=$P_{kom}$ – $P_{true}$ =0

Figure3. Example of KOM algorithm.

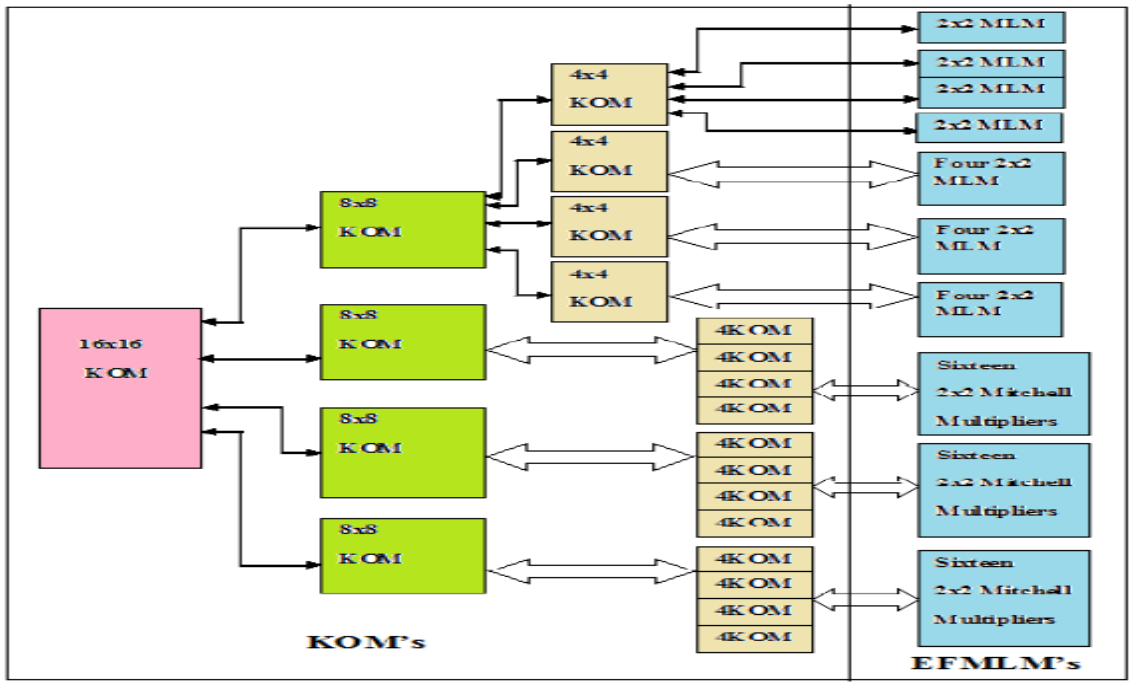

Figure4. Proposed Recursive 16x16 Multiplier.

## 3.1 Proposed Modified 2x2 Error Free MLM(EFMLM)

The MLM produces error product due to approximations made during calculating log and antilog values. The MLM uses piecewise approximation of binary logarithmic curve, which attains actual value at powers of two values while calculating logarithm and antilogarithm values, hence for values having powers of two, errors don't occur. Therefore MLM with any one or both operands with powers of two doesn't produce error which leads in designing error free MLM. The 2x2 MLM has 16 possible multiplication combinations. The 15 combinations have one or both operands having a value of power of 2, hence zero error in products. The only one combination i.e. $11_{(2)}$ x $11_{(2)}$ has non powers of 2, hence error in product occur as shown in Table 1. The correction term is used to eliminate error in MLM to derive EFMLM. The MLM product for $11_{(2)}$ x $11_{(2)}$ is $1000_{(2)}$ instead of actual product value $1001_{(2)}$. The error in product is $1001_{(2)}$ -$1000_{(2)}$ =$0001_{(2)}$. The correction term value of $0001_{(2)}$ is added to MLM product to eliminate error. The flow diagram of 2x2 EFMLM is as shown in Figure 5. The zero detector is used to check values of operands A and B for zero, if any operand is zero then the product is $0000_{(2)}$, else Log values of A and B are computed. The antilog is used on summation of log A and log B to obtain MLM product. The product value is checked for $1000_{(2)}$ and incremented by 1 to correct error in MLM product to obtain true product else MLM product will be the true product.

Table 1. Comparisons of MLMP with RMP for 2 Bit operands

| 2 Bit Operands | | | | Real Multiplier Products( RMP) | | | | Mitchelle Log Multiplier Products (MLMP) | | | | Error Occurance / Position |
|---|---|---|---|---|---|---|---|---|---|---|---|---|
| $A_1$ | $A_0$ | $B_1$ | $B_0$ | $P_3$ | $P_2$ | $P_1$ | $P_0$ | $P_3$ | $P_2$ | $P_1$ | $P_0$ | |
| 0 | 0 | 0 | 0 | 0 | 0 | 0 | 0 | 0 | 0 | 0 | 0 | No |
| 0 | 0 | 0 | 1 | 0 | 0 | 0 | 0 | 0 | 0 | 0 | 0 | No |
| 0 | 0 | 1 | 0 | 0 | 0 | 0 | 0 | 0 | 0 | 0 | 0 | No |
| 0 | 0 | 1 | 1 | 0 | 0 | 0 | 0 | 0 | 0 | 0 | 0 | No |
| 0 | 1 | 0 | 0 | 0 | 0 | 0 | 0 | 0 | 0 | 0 | 0 | No |
| 0 | 1 | 0 | 1 | 0 | 0 | 0 | 1 | 0 | 0 | 0 | 1 | No |
| 0 | 1 | 1 | 0 | 0 | 0 | 1 | 0 | 0 | 0 | 1 | 0 | No |
| 0 | 1 | 1 | 1 | 0 | 0 | 1 | 1 | 0 | 0 | 1 | 1 | No |
| 1 | 0 | 0 | 0 | 0 | 0 | 0 | 0 | 0 | 0 | 0 | 0 | No |
| 1 | 0 | 0 | 1 | 0 | 0 | 1 | 0 | 0 | 0 | 1 | 0 | No |
| 1 | 0 | 1 | 0 | 0 | 1 | 0 | 0 | 0 | 1 | 0 | 0 | No |
| 1 | 0 | 1 | 1 | 0 | 1 | 1 | 0 | 0 | 1 | 1 | 0 | No |
| 1 | 1 | 0 | 0 | 0 | 0 | 0 | 0 | 0 | 0 | 0 | 0 | No |
| 1 | 1 | 0 | 1 | 0 | 0 | 1 | 1 | 0 | 0 | 1 | 1 | No |
| 1 | 1 | 1 | 0 | 0 | 1 | 1 | 0 | 0 | 1 | 1 | 0 | No |
| 1 | 1 | 1 | 1 | 1 | 0 | 0 | 1 | 1 | 0 | 0 | 0 | Yes / LSB of MLMP |

The multiplicand of 2x2 MLM is expressed using equation 20.

$$N1 = \sum_{i=0}^{1} 2^i Z_i$$

(20)

where $Z_i$ is the value of binary number at $i^{th}$ position. Similarly, the multiplier of 2x2 MLM is expressed as given in equation 21.

$$N2 = \sum_{i=0}^{1} 2^i Z_i$$

(21)

The product of 2x2 MLM is expressed as given in equation 22.

$$P_{MLM} = 2^{k_1+k_2}(1+x_1+x_2), \quad for \ x_1+x_2 < 1$$
$$= 2^{k_1+k_2+1}(x_1+x_2), \quad for \ x_1+x_2 \geq 1$$

(22)

The error in multiplication can be eliminated if the following correction terms are added to the $P_{MLM}$.

The Error Free Mitchell log multiplier is given by equation 23.

$$P_{true} = 2^{k_1+k_2}(1+x_1+x_2) + \prod_{i=0}^{1} z_i \quad for \ x_1+x_2 < 1$$
$$= 2^{k_1+k_2+1}(x_1+x_2) + \prod_{i=0}^{1} z_i \quad for \ x_1+x_2 \geq 1$$

(23)

The proposed modified EFMLM eliminates successive multiplication to correct errors as compared to Mitchell log multiplier.

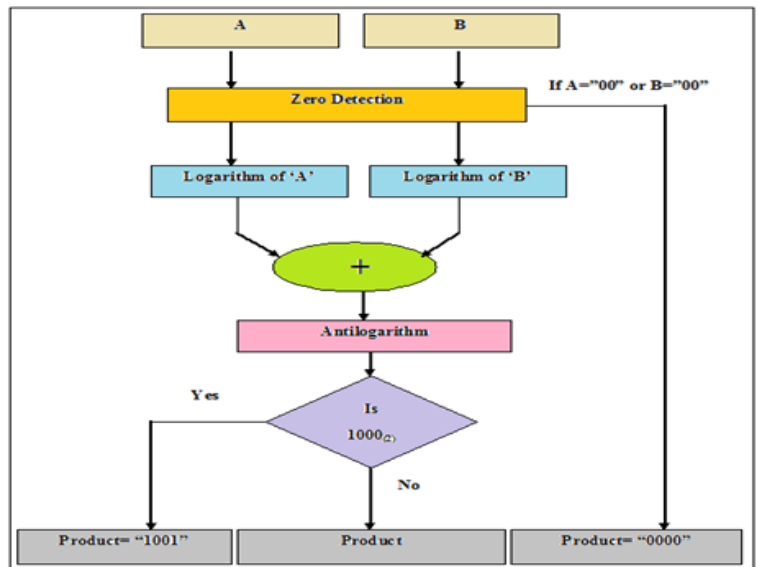

Figure 5. Flow diagram of 2x2 EFMLM

## 3.2 KOM in Proposed Model

The KOM is high speed, parallel multiplier architecture. The product of KOM is given in equation 20.

$$P_{KOM} = a_L b_L + (a_L b_H + a_H b_L) z^{n/2} + (a_H b_H) z^n \qquad (24)$$

Where z=2, for binary number system

The architecture of KOM used in the proposed model is shown in Figure 6. The operands say $a$ and $b$ of size $n$ bits are considered for multiplications. The each $n$ bit operand is decomposed into two $n/2$ bits operands as $a_L, a_H$ of operand $a$ and $b_L b_H$ of operand $b$. The performance of KOM is enhanced using pipelined architecture concept as shown in Figure 7. The architecture is organized into five stages viz., (i) Operand Decomposition stage (ii) Adder 1 stage, (iii) Adder 2 stage, (iv) Adder 3 stage and (v) Operand Alignment stage. With pipelining implementation speed of multiplier is optimized. The Speed of pipelined KOM is double compared to non pipelined KOM.

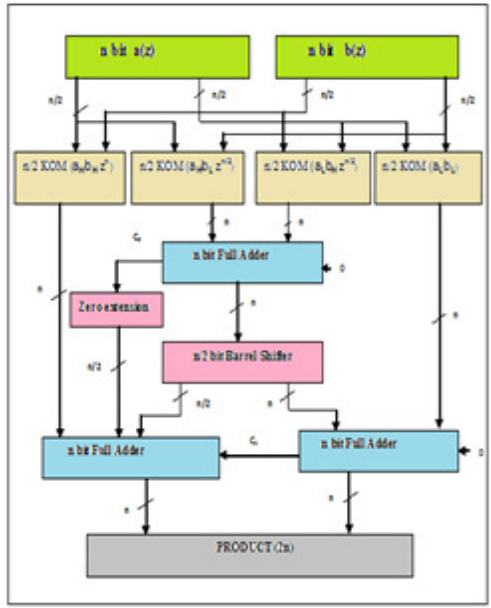

Figure 6. KOM in proposed model

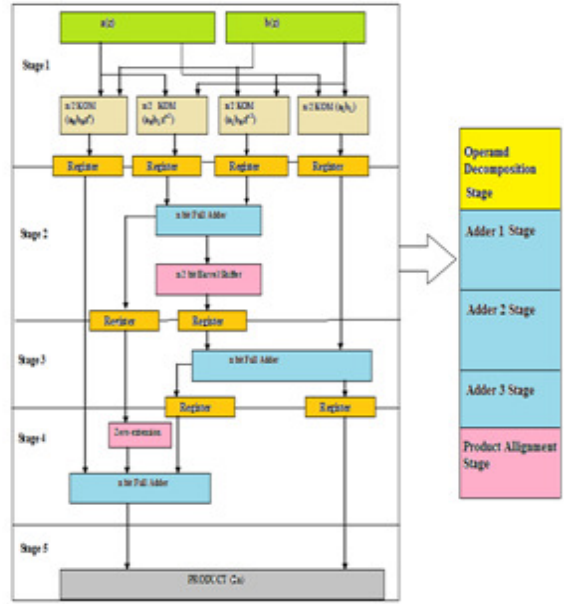

Figure 7. Pipelined Implementation of KOM in proposed model.

### 3.2.1 Throughput analysis of Pipelined KOM.

The pipelined organization for throughput analysis in KOM is shown in Figure 8. To analyze throughput, the one clock period is considered for each operation in pipelined architecture. During first clock period $n$ bit operands are decomposed into four $n/2$ bit operands and are sent to $n/2$ KOM stage. In the second period, the first $n/2$ KOM outputs product, which is Partial Product say pp1 for $n^{th}$ KOM stage. At third clock period, second $n/2$ gives product, which will be pp2 for $n^{th}$ stage. During fourth period, third $n/2$ KOM produces product say pp3, simultaneously pp1 and pp2 are added in adder 1 of $n^{th}$ stage KOM. During fifth clock period, fourth $n/2$ KOM produces pp4, simultaneously pp3 is added with sum of pp1 and pp2 in adder 2 of $n^{th}$ KOM. In sixth clock period the product of fourth $n/2$ KOM pp4 is added with sums of pp1, pp2 and pp3 in adder 3 of $n^{th}$ KOM. In seventh clock period, the product is aligned. In general the pipelined KOM require seven periods to obtain product. The KOM without pipeline require nine clock pulses such as one for operand decomposition, four for partial products generation, three for adders and one for product alignment. As the number of clock periods required by pipelined KOM are less by 2 numbers compared to KOM without pipelined, hence the throughput is enhanced.

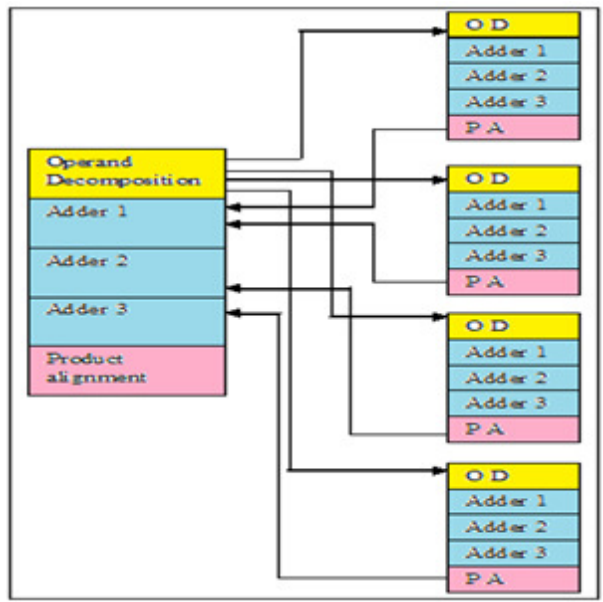

Figure 8. Pipelined organization for throughput analysis in KOM.

### 3.3 Application of Proposed multiplier to fingerprint image smoothening through Gaussian filter.

The Biometrics is used for person identification effectively since biometric traits cannot be shared or misplaced. The several biometric traits such as fingerprint, iris, palmprint, face, voice, signature etc., are used to authenticate a person. The biometric samples may be having some sort of noise with high frequency components, hence preprocessing is required to eliminate high frequency components using filters. The fingerprint database FVC2004 [16], is considered and Gaussian filter along with proposed multiplier is used to eliminate high frequency components in the preprocessing stage of biometric system. In Gaussian filters the fingerprint image is convolved with Gaussian kernel values to eliminate high frequency components. The 2-D, zero mean Gaussian function is given in equation (21) and is sampled and truncated to obtain 3x3 Gaussian kernels as shown in Figure 9 for scaling factor 256.

$$G(x, y) = \frac{1}{2\pi\sigma^2} e^{\frac{-(x^2 + y^2)}{2\sigma^2}} \tag{25}$$

where $\sigma$ is standard deviation of the distribution

$$\frac{1}{256} * \begin{array}{|c|c|c|} \hline 21 & 31 & 21 \\ \hline 31 & 48 & 31 \\ \hline 21 & 31 & 21 \\ \hline \end{array}$$

Figure 9. Gaussian 3x3 Kernel windows for $\sigma$ =1.0 with scaling factor 256.

The 3x3 matrix is considered from the fingerprint image and convolved with 3x3 Gaussian Kernel window using an equation 26. The convolution process is continued for entire image by shifting one column every time.

$$y(m,n) = \sum_{i=1}^{rows} \sum_{j=1}^{columns} h(i,j) \cdot x(m-i, n-j) \qquad (26)$$

where   $x$ is the input fingerprint image.

$h$ is the filter mask.

$y$ is the output image.

*rows* is number of rows in fingerprint image.

*columns* is number of columns in fingerprint image.

### 3.3.1 Hardware Implementation for Fingerprint Image filtering.

The fingerprint image is filtered using Gaussian kernel and proposed multiplier, by adopting convolution and the corresponding architecture is shown in Figure 10. The architecture consists of three First in First out (FIFO) buffers to read first three rows of fingerprint image at every rising edge of clock.

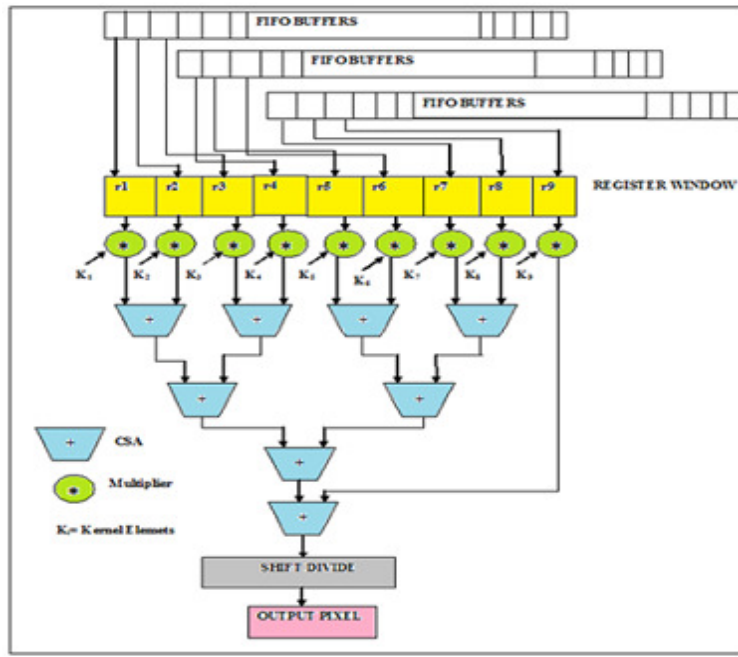

Figure10. Fingerprint filter implementation with proposed multiplier architecture

The first three elements of each buffer are placed in register window to constitute 3x3 Finger print image matrix. The Gaussian Kernel window of size 3x3 is convolved with 3x3 register window using proposed multiplier and Carry Save Adders (CSA). The register window of fingerprint image is shifted right by one column and convolved with Gaussian Kernel and this process is continued till the end column of image to complete process for first three  rows of an image. The rows are shifted down by one to consider second, third and fourth rows for convolution process and are continued till end row of the fingerprint image to complete convolution process with whole fingerprint image.

## 4. ALGORITHM

*Problem Definition:* The higher order efficient error free multiplier is designed by combining 2x2 REFMLM and KOM. The Proposed multiplier is used to convolve finger print image with Gaussian Filter kernel window of size 3x3, to enhance image quality for fingerprint applications.

The objectives are:
1. Design of error free higher order multiplier.
2. To use proposed multiplier in Gaussian smoothing filter for noise removal in finger print images, thereby increasing Peak Signal to Noise Ratio (PSNR) of image.

The proposed algorithm is product implementation of REFMLM and KOM as given in Table 2. The KOM product equations are given in equation 27.

$$P_{KOM} = a_L b_L + (a_L b_H + a_H b_L) z^{n/2} + (a_H b_H) z^n$$
$$= low + (mid1 + mid2) z^{n/2} + (high) z^n \tag{27}$$

Where Low= $a_L b_L$, Mid1= $a_H b_L$, Mid2= $a_L b_H$ and High= $a_H b_H$. The equation 23 is used to decompose $n$ x $n$ multiplier till 2x2 stages. The REFMLM is designed for 2 x 2 stage with error correction term.

#### Table 2. Proposed Multiplier Algorithm

| Inputs: *operands a and b of size n bits with powers of 2.* Output: *Product P of size 2n bits* |
|---|
| Step1: $a_L$=a[0 to (n/2-1)]; ←Decompose $a$ into Lower $n/2$ size operand $a_L$. |
| Step2: $a_H$=a[(n/2) to (n-1)] ; ← Decompose $a$ into Higher $n/2$ size operand $a_H$ |
| Step3: $b_L$=b[0 to (n/2-1)]; ←Decompose $b$ into Lower $n/2$ size operand $b_L$ |
| Step4: $b_H$=b[(n/2) to (n-1)] ; ←Decompose $b$ into Higher $n/2$ size operand $b_H$ |
| Step5: low= KOM ($a_L b_L$); ← operands sent to $n/2^{th}$ KOM stage to get low |
| Step6: high= KOM ($a_H b_H$); ←operands sent to $n/2^{th}$ KOM stage to get high |
| Step7: mid1=KOM ($a_H b_L$); ← operands sent *to* $n/2^{th}$ KOM stage to get mid1 |
| Step8: mid=KOM ($a_L b_H$); ← operands sent to $n/2^{th}$ KOM stage to get mid2 |
| Step 9: The $n/2^{th}$ stage is decomposed into $n/4$, $n/8$….till 2x2, by repeating steps 1 to 8. |
| Step 10: Calculate Characteristic number $k_1$; ←leading one position of *first* 2 bit operand of 2x2 stage. |
| Step 11: Calculate Characteristic number $k_2$; ← leading one position of *second* 2 bit operand of 2x2 stage. |
| Step 12: Calculate $d_1$← Decoded value of $k_1$ in binary. |
| Step 13: Calculate $d_2$← Decoded value of $k_2$ in binary. |
| Step 14: Calculate mantissa $x_1$; ← shift (*First* 2 bit operand - $d_1$) to the left by $k_2$ bits |
| Step 15: Calculate mantissa $x_2$; ← shift (*Second* 2 bit operand − $d_2$) to the left by $k_1$ bits |
| Step 16: Calculate $k_{12}$= $k_1$ + $k_2$ ←Characteristics sum |
| Step 17: Calcultae $d_{12}$← Decoded value of $k_{12}$ in binary. |
| Step 18: Calculate $x_{12}$= $x_1$ + $x_2$ ←mantissas sum |
| Step 19 $P_{MLM}$= $d_{12}$+$x_{12}$; ←Approximated Mitchelle Log Multiplier Product |
| Step 20: $P_{REFMLM}$=1001    if First operand and second operand =$11_{(2)}$ else $P_{REFMLM}$=$P_{MLM}$. |
| Step 21: The four $P_{REFMLM}$ 's of 2x2 stage constitute Partial products for 4x4 stage. |
| Step 22: mid= mid1+mid2;← mid1 and mid2 partial products added to get mid. |
| Step 23: $P_{KOM4x4}$=high+mid+low;←Product of 4x4 stage is obtained. |
| Step 24: Products of 4x4 stage constitute Partial products of 8x8 stage. |
| Step 25: Steps 22 and 24 are repeated for higher stages of KOM, till $n$ x $n$ Multiplier stage |

# 5. DEFINITIONS OF PERFORMANCE PARAMETERS

### 5.1.1 Error Rate (ER)

The percentage of error present in logarithmic multiplier and is measured with equation 28

$$\% \, ER = \frac{P_{true} - P_{error}}{P_{true}} * 100\%$$

(28)

Where $P_{true}$ - the true product of multiplication,

$P_{error}$- error product obtained from Log multiplier.

The maximum value of ER is called Maximum Error Rate (MER).

### 5.1.2 Average Error Rate(AER)

The mean value of error between true and error product is AER given in equation 29.

$$AER = \left( \frac{1}{n} \sum_{i=0}^{n} error_i \right) * 100\%$$

(29)

Where $error = P_{true} - P_{error}$.

$n=$ Number of multiplications carried.

### 5.1.3 Mean Square Error (MSE)

The cumulative squared error between the corrupted and the original image and is given in equation 30.

$$MSE = \frac{\sum_{i=1}^{M} \sum_{j=1}^{N} [I_b(i,j) - I_c(i,j)]^2}{M * N}$$

(30)

Here $I_b(i, j)$ = Base image,

$I_c(i, j)$ =Corrupted Image,

$M$ = number of rows in an image and

$N$ = number of columns in an image.

### 5.1.4 Peak Signal to Noise Ratio(PSNR)

The Signal with respect to noise between corrupted and the original image and is measured using PSNR equation 31.

$$PSNR = 10 \log_{10} \left( \frac{R^2}{MSE} \right)$$

(31)

*Here R* is the maximum fluctuation in the input image data type. For example, if the input image has a double-precision floating-point data type, then *R* is 1. If it has an 8-bit unsigned integer data type, *R* is 255.

## 5.2 Performance Results

### 5.2.1 Device Utilisation

All designs referred are implemented using Very High Speed Integrated Circuit (VHSIC) Hardware Description Language (VHDL). Multipliers are designed and implemented on Spartan 3 family device, XC3S1500-5fg320. CAD tool used for design implementation in Xilinx ISE 9.2i release. The VHDL implemented proposed multiplier is simulated using Modelsim 5.7g and then place-and-routed using Xilinx ISE.

Table3. Comparison of area utilizations for existing and proposed non pipelined multipliers.

| Multiplier (16x16) | Slices | 4 input LUT'S | IOB'S |
|---|---|---|---|
| MA[18] | 321 | 622 | 99 |
| OD-MA[18] | 604 | 1187 | 99 |
| BB [18] | 276 | 533 | 99 |
| BB+1ECC [18] | 564 | 1099 | 99 |
| BB+2ECC [18] | 814 | 1596 | 99 |

Table 4. Comparison of area utilizations for existing and proposed pipelined multipliers

| Multiplier (16x16) | Slices | 4 input LUT'S | IOB'S |
|---|---|---|---|
| BB [18] | 216 | 404 | 99 |
| BB+1ECC [18] | 427 | 803 | 99 |
| BB+2ECC [18] | 635 | 1189 | 99 |
| BB+3ECC [18] | 824 | 1546 | 99 |
| Proposed | 662 | 1106 | 65 |

The area utilizations in terms of slices, LUT's and IOB's of proposed non pipelined and pipelined multipliers is compared with existing algorithm[18] are given in Table 3 and Table 4 respectively. The three stage error correction term in existing MLM [18] increases the area utilization. It's clearly observable that proposed multiplier need less area compared to existing log multiplier with 3 ECC since it uses 2x2 error free MLM. The error free 2x2 MLM is used recursively in KOM to derive high order multipliers. The area utilization is reduced in proposed method compared to existing method since only one error correction term is required in base MLM.

### 5.2.2 Operating frequency (Speed)

Operating frequency of the device is a measure of how fast the output is obtained for given input combinations. Operating frequency of existing multipliers and proposed multipliers are compared in Table 5. The proposed multiplier operating speed is high with non pipelined architecture. The operating speed of proposed multiplier with pipelined multiplier is less in case of proposed method compared to existing method since there exists clock latency between stages of multipliers.

Table5. Operating frequency comparisons of proposed and existing multipliers

| Multiplier | Non-pipelined(MHz) | Pipelined(MHz) |
|---|---|---|
| MA [18] | 50.554 | - |
| OD-MA [18] | 40.330 | - |
| BB [18] | 58.075 | 153.335 |
| BB+1ECC [18] | 50.180 | 153.335 |
| BB+2ECC [18] | 41.429 | 153.335 |
| BB+3ECC [18] | 37.826 | 153.335 |
| PROPOSED | 66.909 | 130.511 |

### 5.2.3 Multiplier Error

The AER and MER values of proposed architecture are compared with existing architecture for 16x16 multipliers are given in Table 6

Table6. Error comparisons of proposed and existing multipliers.

| Multiplier (16X16) | AER | MER |
|---|---|---|
| MA [18] | 3.82% | 11.11% |
| OD-MA [18] | 3.53% | 11.11% |
| BB[18] | 9.41% | 25% |
| BB+1CT [18] | 0.98% | 6.25% |
| BB+2CT [18] | 0.11% | 1.56% |
| BB+3CT [18] | 0.01% | 0.39% |
| Proposed(with Error Correction) | 0.00% | 0.00% |

The AER value is high for Mitchell Algorithm (MA) i.e., 3.82% as against AER value of 0.00% in proposed multiplier. The MER value is high for Mitchelle Algorithm (MA) i.e., 11.11% as against MER value of 0.00% in the proposed multiplier. It is observed that the AER and MER values are 0.00% compared to existing multipliers having the values more than zero since the error in basic 2x2 multiplier is already corrected, hence error is not propagated to higher order KOM architecture.

### 5.2.4 RTL Schematics

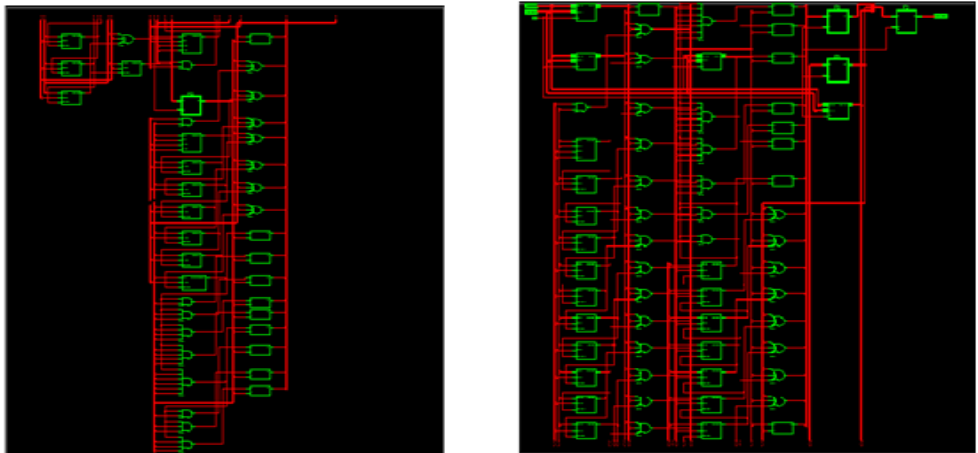

Figure 11. RTL schematics of proposed multiplier on XC3S1500-5fg320 FPGA.

The Register transfer Level of Proposed EFMLM is shown in Fig 11. The 8x8 EFMLM is targeted on Spartan 3 FPGA family device XC3S1500-5fg320. It is observed that the proposed architecture consumes only 912 LUT's for 16x16 Non-Pipelined multiplier.

### 5.2.4b Comparison of 4x4 REFMLM multiplier using different parameters.

The proposed 4x4 multiplier is compared with existing multipliers for different performance parameters are given in Table 7. The zero error is achieved through BB+3ECC with 117 LUTs, but the proposed technique with error correction term achieves same zero error with only 33 LUTs

Table 7. Comparisons of proposed 4x4 REFMLM with existing multipliers.

| Parameters | LUT's | % AER | % MER | Maximum Combinational Path Delay (ns) |
|---|---|---|---|---|
| Mitchell[18] | 53 | 5.5185 | 11.11 | 17.638 |
| ODMA[18] | 124 | 3.58515 | 7.2453 | 26.499 |
| BB[18] | 50 | 5.05925 | 21.77 | 12.800 |
| BB+1ECC[18] | 88 | 0.28889 | 4 | 16.221 |
| BB+2ECC[18] | 109 | 0.0074007 | 0.4444 | 18.737 |
| BB+3ECC[18] | 117 | 0.000000 | 0.000 | 19.083 |
| Proposed Without Error Correction | 33 | 1.7629 | 11.111 | 16.603 |
| Proposed with Error Correction | 33 | 0.00000 | 0.0000 | 16.638 |

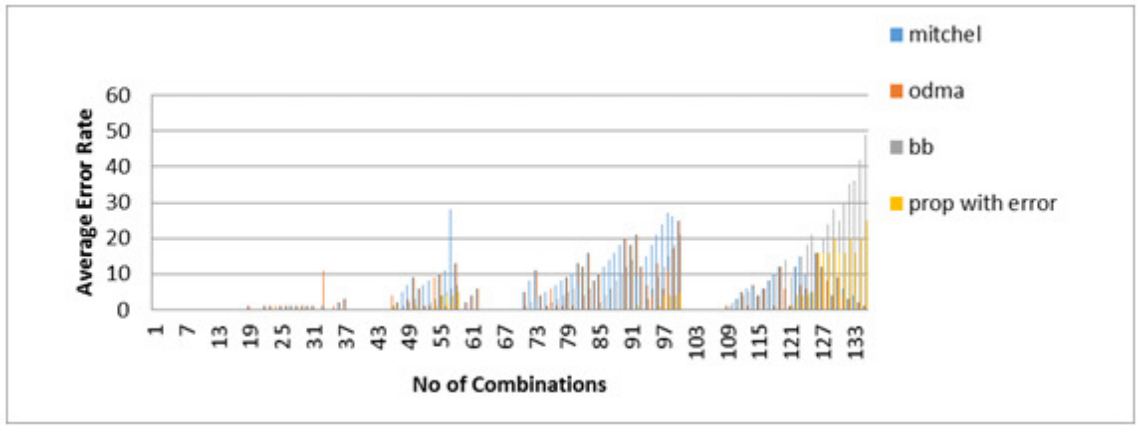

Figure12. Comparison of AER for Proposed Mitchell, ODMA and Basic Block

The delay of the proposed multiplier is 16.638 ns compared to 19.083 ns in case of BB+3ECC. If an error is allowed to propagate in 2x2 mitchell to build 4x4 multiplier, the AER of 1.7629% is obtained as shown in Table 7. The graph has been plotted to show AER values of proposed 4x4 multiplier for 134 unique combinations of multiplications is shown in Figure 12. The value of AER is less in the proposed multiplier with error propagation compared to mitchell, ODMA and BB.

The Table 8 shows the Relative error rate of proposed with and without error correction terms of 2x2 multiplier propagated to 4x4 multiplier. The different values of relative error rates are compared with BB, BB+1ECC and BB+2ECC.

Table 8. Comparison of relative error rates of proposed 4x4 multiplier with different existing multipliers.

| Error Rate (%) | Percentage of 4x4 multiplier combinations. | | | | |
|---|---|---|---|---|---|
| | BB | BB+1ECC | BB+2ECC | WITH ERROR | WITHOUT ERROR/BB3 ECC |
| 0.00 | 51.47 | 88.20 | 99.26 | 79.4 | 100 |
| 0.0>ER<0.05 | 6.6 | 11.7 | 0.73 | 8 | 0.0 |
| 0.05>ER<0.1 | 24.26 | 11.7 | 0.73 | 13.2 | 0.0 |
| 0.1>ER<0.5 | 48.52 | 11.7 | 0.73 | 20.5 | 0.0 |
| 0.5>ER<1.0 | 48.52 | 11.7 | 0.73 | 20.5 | 0.0 |

### 5.2.5 4x4 Multiplier extended to 16x16 multiplier.

The comparisons of 4x4 error free BB+3ECC, Iterative BB+3ECC and proposed error correction when extended to 16x16 multipliers are shown in Table 9. The proposed error correction will utilize 733 LUTs since the BB+3ECC method requires repetitive error correction circuits.

Table 9. Comparison of proposed 16x16 multiplier with error free BB+3ECC 16x16 multiplier

| Parameters | Error free BB+3ECC extended to 16x16(KOM) | BB+3ECC Iterative Multiplier | Proposed with EC 16x16 |
|---|---|---|---|
| Hardware Utilizations (Number of 4 input LUT's) | 1888 | 1438 | 733 |
| Maximum Combinational Path Delay (ns) | 61.181 | 41.555 | 53.837 |

### 5.2.6 Noise Reduction in fingerprint image using proposed multiplier

The noise in fingerprint image is random and is generated by scanners. The noise in image may be of Gaussian noise, Salt and Pepper noise, shot noise, Quantization noise etc. The fingerprint image is considered from FVC2004 database for noise analysis. The salt and Pepper noise is added to the original fingerprint image. The Gaussian filters are used to eliminate salt and pepper noise in fingerprint image with ODMA, Iterative Log Multiplier (3ECC) and proposed multiplier shown in Figure 13.

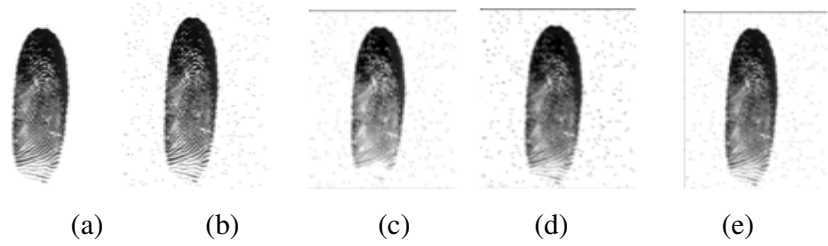

(a)          (b)          (c)          (d)          (e)

(a) Original Fingerprint, (b) Salt and pepper noise added to base image, (c) Smoothened image using ODMA multiplier in filter, (d) Smoothened image using pipelined log multiplier with 3ECC in filter, (e) Smoothened image using proposed multiplier in filter.

Figure 13. Gaussian Smoothening to remove noise in fingerprint image using different multipliers.

The salt and pepper added to the original fingerprint is eliminated effectively using proposed multiplier compared to existing multilpliers such as ODMA and BB+3ECC in Gaussian smoothing. The PSNR values of corrupted image with noise and smoothened images are tabulated in Table 10 for scaling factors of Gaussian kernels 256. The percentage Salt and pepper noise levels are varied from 10 to 40. The PSNR values decreases as percentage of noise level increases. It is observed that the proposed multiplier has better PSNR compared to existing multipliers for all noise levels, since the 8x8 multiplier designed is error free which is corrected using error correction circuit.

Table 10. Comparisions of PSNR values for proposed and existing multipliers with Gaussian filter

| noise (%) | multipliers | PSNR of corrupted image(dB) | PSNR of Smoothend image(dB) |
|---|---|---|---|
| 10 | ODMA [19] | 13.4245 | 6.3938 |
| | Iterative 3ECC [18] | 13.4245 | 5.2479 |
| | Proposed | 13.4245 | 15.2429 |
| 20 | ODMA [19] | 10.3667 | 5.9971 |
| | Iterative 3ECC [18] | 10.3667 | 4.9155 |
| | Proposed | 10.3667 | 13.9784 |
| 30 | ODMA [19] | 8.6004 | 5.6198 |
| | Iterative 3ECC [18] | 8.6004 | 5.1197 |
| | Proposed | 8.6004 | 12.6906 |
| 40 | ODMA [19] | 7.3684 | 5.2732 |
| | Iterative 3ECC [18] | 7.3684 | 5.4866 |
| | Proposed | 7.3684 | 11.5341 |

## 6. CONCLUSIONS

In this paper, we have proposed a new approach to derive Recursive Error Free Mitchell Log Multiplier (REFMLM), which is used in KOM for digital image processing applications. The proposed multiplier is combination of 2x2 EFMLM with parallel KOM architecture. The EFMLM is based on introducing correction term for 2x2 Mitchell log multiplier when both multiplicand and multiplier has binary value 11(2). The 2x2 EFMLM is used as basic multiplier block for KOM and extended upto 16x16 multiplier. The multiplier is used in convolution of noisy fingerprint image with Gaussian mask to derive filtered fingerprint image to enhance the quality of fingerprint image for biometric identification. The performance parameters such as area utilization, speed, error and PSNR are better in the case of proposed architecture compared to existing architectures. In Future the 4x4 error free multiplier can be designed and extended to higher order multipliers.

**AUTHORS**

**Satish S Bhairannawar** is an Associate Professor, Dept of Electronics & Communication, Dayananda Sagar College of Engineering, Bangalore. He obtained his B.E degree in Electronics & Communication from Bangalore University and M.E. Degree in Electronics and communication from University Visvesvaraya College of Engineering, Bangalore. He is pursuing his Ph.D. in Computer Science and Engineering at Bangalore University, Karnataka. His research interests include Image Processing, Biometrics, VLSI for signal processing applications, Video Processing, VLSI Design and Circuits.

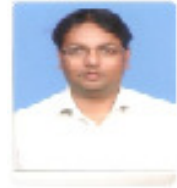

**Rathan R** is an Design Engineer in R&D Department of System Controls Technology Solutions Pvt. Ltd. He obtained his B.E from Bangalore University and M.E in Electronics and Communication Engineering from University Visvesvaraya College of Engineering, Bangalore. His research interests include Image Processing, Efficient hardware architectures for signal and image filters, Embedded Image Hardware designs.

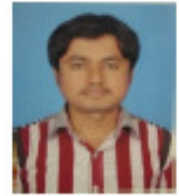

**K B Raja** is an Assistant Professor, Dept. of Electronics and Communication Engineering, University Visvesvaraya college of Engineering, Bangalore University, Bangalore. He obtained his B.E and M.E in Electronics and Communication Engineering from University Visvesvaraya College of Engineering, Bangalore. He was awarded Ph.D. in Computer Science and Engineering from Bangalore University. He has over 110 research publications in refereed International Journals and Conference Proceedings. His research interests include Image Processing, Biometrics, VLSI Signal Processing, computer networks.

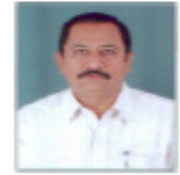

**K R Venugopal** is currently the Principal and Dean, Faculty of Engineering, University Visvesvaraya College of Engineering, Bangalore University, Bangalore. He obtained his Bachelor of Engineering from University Visvesvaraya College of Engineering. He received his Masters degree in Computer Science and Automation from Indian Institute of Science, Bangalore. He was awarded Ph.D. in Economics from Bangalore University and Ph.D. in Computer Science from Indian Institute of Technology, Madras. He has a distinguished academic career and has degrees in Electronics, Economics, Law, Business Finance, Public Relations, Communications, Industrial Relations, Computer Science and Journalism. He has authored 27 books on Computer Science and Economics, which include Petrodollar and the World Economy, C Aptitude, Mastering C, Microprocessor Programming, Mastering C++ etc. He has been serving as the Professor and Chairman, Department of Computer Science and Engineering, University Visvesvaraya College of Engineering, Bangalore University, Bangalore. During his three decades of service at UVCE he has over 350 research papers to his credit. His research interests include computer networks, parallel and distributed systems, digital signal processing and data mining.

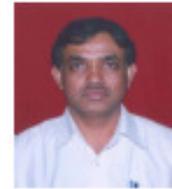

**L M Patnaik** is the Honorary Professor, Indian Institute of Science, Bangalore, India. During the past 35 years of his service at the Indian Institute of Science, Bangalore, He has over 600 research publications in refereed International Journals and Conference Proceedings. He is a Fellow of all the four leading Science and Engineering Academies in India; Fellow of the IEEE and the Academy of Science for the Developing World. He has received twenty national and international awards; notable among them is the IEEE Technical Achievement Award for his significant contributions to high performance computing and soft computing. His areas of research interest have been parallel and distributed computing, mobile computing, CAD for VLSI circuits, soft computing, and computational neuroscience.

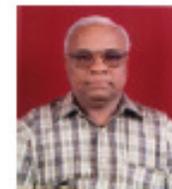